\documentclass[reprint,amsmath,amssymb,aps,showkeys]{revtex4-2}

\usepackage{slashed}
\usepackage{amsmath}

\usepackage{graphicx}
\usepackage{dcolumn}
\usepackage{bm}
\usepackage{hyperref}
\usepackage[mathlines]{lineno}
\usepackage{xcolor}
\usepackage{upgreek}
\usepackage[makeroom]{cancel}
\usepackage{upgreek}

\newcommand\RR{\mathbb{R}}
\newcommand\CC{\mathbb{C}}

\newcommand\id{\textit{id}}

\renewcommand\H{\mathcal{H}}

\renewcommand\L{\mathcal{L}}

\newcommand\SU{\mathcal{SU}}
\newcommand\U{\mathcal{U}}
\newcommand\SO{\mathcal{SO}}

\renewcommand\O{\mathcal{O}}

\newcommand\vphi{\varphi}

\newcommand\rarrow{\rightarrow}

\newcommand\diff{\mathfrak{diff}}

\newcommand\gl{\mathfrak{gl}}

\renewcommand\t{\tilde}

\renewcommand\b{\bar }

\renewcommand\d{\partial}

\newcommand\bs{\boldsymbol}

\renewcommand\-{^{-1}}

\renewcommand\id{\text{id}}

\DeclareMathOperator{\Diff}{Diff}

\begin{document}


\title{Spacetime boundaries do not break diffeomorphism and gauge symmetries} 

\author{J. François}
\email{jordan.francois@uni-graz.at}
\affiliation{University of Graz (Uni Graz), 
Heinrichstraße 26/5, 8010 Graz, Austria, and \\
Masaryk University (MUNI), 
Kotlářská 267/2, Veveří, Brno, Czech Republic, and \\
Mons University (UMONS), 
20 Place du Parc, 7000 Mons, Belgium. 
}

\author{L. Ravera}
\email{lucrezia.ravera@polito.it}
\affiliation{Politecnico di Torino (PoliTo),
C.so Duca degli Abruzzi 24, 10129 Torino, Italy, and \\
Istituto Nazionale di Fisica Nucleare (INFN), Section of Torino,
Via P. Giuria 1, 10125 Torino, Italy, and \\
Grupo de Investigación en Física Teórica,
Universidad Cat\'{o}lica De La Sant\'{i}sima Concepci\'{o}n, Chile.
}

\date{\today}

\begin{abstract}

In General Relativity and gauge field theory, one often encounters a claim, which may be called  the 
\emph{boundary problem}, according to which ``boundaries break diffeomorphism and gauge symmetries".
We~argue that this statement has the same conceptual structure as the \emph{hole argument}, and is thus likewise defused by the \emph{point-coincidence argument}: 
We show that the boundary problem dissolves once it is understood that a \emph{physical region}, thus its boundary, is \emph{relationally} and \emph{invariantly defined}. 
This insight can be technically implemented via the Dressing Field Method, a systematic tool to exhibit the gauge-invariant content of general-relativistic gauge field theories, whereby physical field-theoretical degrees of freedom co-define each other and define, coordinatize, the \emph{physical spacetime}. 
We illustrate our claim with a simple application to the case of General Relativity.

\end{abstract}


\maketitle


\section{Introduction}
\label{Introduction}

General-relativistic (GR) physics, which is based on diffeomorphisms, and gauge field theory (GFT), based on gauge symmetries, form together the broad framework of general-relativistic gauge field theory (gRGFT).
A~model within this framework is a field theory on an $n$-dimensional manifold $M$, whose field content we  denote by $\upphi$ and which is acted upon by the covariance group $\Diff(M)\ltimes \H$
{as $\upphi\mapsto\psi^*(\upphi^\gamma)$}, with $\Diff(M)$ the diffeomorphisms group 
and 
{$\H:= \left\{ \gamma: M\rarrow H\, |\, {\gamma_1}^{\!\gamma_2}=\gamma_2\- \gamma_1 \gamma_2\right\}$ the gauge group, and $H$ a Lie group}; 
the field equations for $\upphi$ are covariant under $\Diff(M)\ltimes \H$, hence its name.
Considering such a theory over a bounded region $U\subset M$ with boundary $\d U$,  
one frequently encounters in the  
literature the claim that ``diffeomorphism and/or gauge symmetries are broken at spacetime boundaries" \cite{DonnellyFreidel2016, Speranza2018,Geiller2018,Freidel-et-al2020-1,Kabel-Wieland2022,Ciambelli2023}. 
We may refer to it as the \emph{boundary problem}.
More accurately, what is actually meant  is that ``field configurations, or functional thereof of special interest, defined at $\partial U$ are not invariant under the action of $\Diff(M)\ltimes \H$."
The field configurations in question may be some choice of gauge, the functional of the field at $\d U$ may be the presymplectic potential or 2-form of the theory, objects of interest~notably~in~the {covariant} phase space analysis of the symplectic structure of gRGFTs \cite{Witten1986,Zuckerman1986,Crnkovic1987, CrnkovicWitten1986}, see  \cite{Gieres2021}.

Depending on the context and aims, various counter-measures are put forth to solve the boundary problem.
In the covariant phase space literature, this
involves e.g. the \emph{ad hoc} introduction of so-called \emph{edge modes}, 
 \mbox{degrees} of freedom (d.o.f.) typically confined to $\partial U$, 
whose gauge transformations are tuned to cancel 
the terms from the transformation of the symplectic potential and 2-form, thereby  ``restoring their invariance" \cite{DonnellyFreidel2016, Speranza2022,Geiller2017,Speranza2018, Geiller2018,Chandrasekaran_Speranza2021,Kabel-Wieland2022}.
Edge~modes are sometimes interpreted as ``Goldstone modes" resulting from the breaking of $\Diff(M)\ltimes \H$~at~$\d U$. 
Some have  claimed that edge modes reveal essential new symmetries of gravity (so-called ``corner symmetries") and may be key to a new paths to quantum gravity (QG) \cite{Freidel-et-al2020-1, Freidel-et-al2020-2, Freidel-et-al2020-3,Ciambelli2023}. 
\smallskip

However, we show that there is no boundary \mbox{problem}. 
Indeed, an analysis from first principles reveals  it to be logically equivalent to 
Einstein's famous \emph{hole \mbox{argument}} 
in GR \cite{Stachel1986, Norton1987,Norton1988,Stachel1989,EarmanNorton,Norton1987,Norton1993,Stachel2014}, and its generalization to GFT and gRGFT \cite{Stachel2014, JTF-Ravera2024c,BFR-fbsubst}. 
The hole argument was designed to highlight an apparent ill-posedness of the Cauchy problem, and attending breakdown of determinism, in $\Diff(M)$-covariant theories. Its generalization shows \mbox{likewise}~in $\H$-covariant theories.
Einstein only \mbox{successfully} completed GR 
\mbox{after} he resolved this dual conceptual/technical \mbox{issue} via
what came to be known, after Stachel named it thus, as the \emph{point-coincidence argument} -- see  \cite{Stachel1986, Stachel1989, Stachel2014, Giovanelli2021, JTF-Ravera2024c} and references therein.
A generalized point-coincidence argument  establishes  
\mbox{\emph{relationality}}  at the paradigmatic \mbox{conceptual} core of gRGFT: 
namely the fact that \mbox{physical} magnitudes, i.e. 
physical field-theoretical d.o.f., \mbox{relationally} co-define each other,
as well as \mbox{physical} 
 \emph{events}, 
in a $\big(\!\Diff(M)\ltimes\H\big)$-invariant~way. 

This insight ought to defuse the boundary problem. 
That it did not is owed to the fact that the \mbox{relationality} of gRGFT is only \emph{tacit} in its standard formalism, \mbox{encoded} in its \emph{manifest covariance} under the group $\Diff(M)\ltimes \H$. 
This makes the physics of gRGFT sometimes tricky to read-off from the formalism, and in particular the \mbox{identification} of its observables non-trivial. 


A framework allowing a manifestly relational $\big(\!\Diff(M)\ltimes\H\big)$-invariant reformulation of gRGFTs would make their conceptual structure  transparent, giving ready access to their observables and dissolving the boundary problem. 
Here we shall present such a framework, based on a 
systematic approach to reduce symmetries:
the \emph{Dressing Field Method} (DFM) 
\cite{Francois2023-a,JTF-Ravera2024gRGFT,JTF-Ravera2024-SUSY,JTF-Ravera2024ususyDFM,JTF-Ravera2024NRrelQM,JTF-Ravera2025DFMSusyReview&Misu,JTF-Ravera2025offshellsusy}. 
We will show how it technically implements the point-coincidence argument \cite{BFR-fbsubst} via the definition of relational invariant dressed variables, living on relationally and invariantly defined dressed regions, so that the boundary problem cannot get a grip.

The remainder of this paper is structured as follows: 
In Section \ref{The hole and point-coincidence arguments in a nutshell} we give a dense overview of the conceptual insights stemming from articulating the hole and point-coincidence arguments. 
In Section \ref{Dressing Field Method: basics} we recall the~\mbox{basics} of the DFM for gauge symmetries and diffeomorphisms, 
introducing 
in particular the notion of \emph{dressed regions}.
It is put to use in
Section \ref{Dissolving the boundary problem in General Relativity}, where invariant relational \emph{physical spacetime boundaries} are defined, the boundary problem being thereby averted.
Our~presentation will emphasize for concreteness the explicit example of~GR. 
It will be structured in decreasing levels of abstraction, a progression designed to make the content accessible 
to a broad audience, versed in either differential geometry or field-theoretical/components formulations. 
We gather our closing remarks  
in Section \ref{Conclusions}.

\section{The hole and point-coincidence arguments in a nutshell}
\label{The hole and point-coincidence arguments in a nutshell}

Einstein's dual
\emph{hole argument} and \emph{point-coincidence argument} were key to his final  understanding of the meaning of diffeomorphism covariance leading to the  completion of GR, and to his definite views on spacetime -- magnificently expressed in the last paragraphs of his appendix to \cite{Einstein1952-EINRTS-3}. 
It is an unfortunate turn of {history} that subsequently most of the physics community either forgot about them, or misunderstood them as mere distracting philosophical musings diverting Einstein from the straight path to GR.   
Their fundamental significance was rediscovered and brought to a wider attention by Stachel in 1980 (published in \cite{Stachel1989}), see  \cite{EarmanNorton}-\cite{Norton1987}, and \cite{JTF-Ravera2024c} for an in-depth modernized account.
The literature on the subject is now substantive, even involving parts of the quantum gravity community.
Yet it remains misappreciated by most, which has non-trivial consequences.

The core logic of the hole argument goes as follows.
By assumption, the field equations $ E (\upphi)=0$ of a general-relativistic theory are  $\Diff(M)$-covariant, which implies that if $\upphi$ is a solution, so is $\psi^* \upphi$ for any $\psi \in\Diff(M)$, since $E(\psi^*\upphi)=\psi^* E(\upphi)=0$.
Now, let us
consider two solutions $\upphi, \upphi'$ belonging to the same $\Diff(M)$-orbit $\O_\upphi$, i.e. $\upphi'=\psi^*\upphi$, for which $\psi$ is a compactly supported diffeomorphism whose support $D_\psi \subset M$ is the ``hole": we have that $\upphi=\upphi'$ on $M/D_\psi$, but $\upphi \neq \upphi'$ on $D_\psi$. 
Therefore, the field equations have an ill-defined Cauchy problem, they cannot uniquely determine the evolution of the fields $\upphi$ within $D_\psi$, so that the theory \emph{prima \mbox{facie}} seems to suffer from indeterminism. 
The question then arises; how are deterministically evolving physical spatio-temporal d.o.f. 
represented in GR?

The key conceptual insight came from what Stachel called Einstein's \emph{point-coincidence argument}. 
It \mbox{consists} in 
 the observation that the physical content of the \mbox{theory} (in particular its possible
verifications, and observables) is exhausted by the pointwise coincidental values of fields $\upphi$, 
and that the description of such coincidences is $\Diff(M)$-invariant. 
It means that one has to admit that all solutions within the same $\Diff(M)$-orbit $\O_\upphi$ represent the \emph{same} physical state. 
 The deterministically evolving physical d.o.f. are thus encoded in \emph{equivalence classes} $[\upphi]$ under $\Diff(M)$: 
 they are not instantiated within the individual mathematical fields $\upphi$, but by the $\Diff(M)$-invariant \emph{relations} among them.
 
A further immediate consequence is that the theory, unable to physically distinguish between $\Diff(M)$-related solutions of $ E(\upphi)=0$, consequently cannot  distinguish $\Diff(M)$-related points of $M$ either. 
In other words, points of $M$ are not \emph{physical spatio-temporal events}. 
How, then, is physical spacetime  
represented in GR?
The point-coincidence argument suggested the~\mbox{answer}: 
Points of the physical spacetime being \emph{defined}, \mbox{individuated}, 
as pointwise coincidences of distinct
physical field  d.o.f., the physical spacetime is thus represented in GR as the $\Diff(M)$-invariant \emph{network of relations} among physical field. 
The manifold $M$, like a scaffold to a building, is only there to
bootstrap our ability to build a description of the relativistic physics of fields' spatio-temporal d.o.f.  
And like
a scaffold, $M$ is removed from the physical picture by the  $\Diff(M)$-covariance of the 
 general-relativistic field equations. 
As Einstein concluded \cite{Einstein1952-EINRTS-3}, ``\emph{Physical [fields]  are not in space, but [$\ldots$]
are spatially extended}" so that
``\emph{Space-time does not claim existence on its own, but only as a structural quality of the field}." 
\smallskip

The \emph{internal  hole argument} has a similar \mbox{structure}, \emph{mutatis mutandis}:
The fields $\upphi$ of a GFT have both spatio-temporal and \emph{internal} d.o.f., 
and the $\H$-covariance of their \mbox{field equations}, 
$E(\upphi^\gamma)=\uprho(\gamma)\- E(\upphi)=0$ for any $\gamma \in\H$ -- with
 $\uprho$ denoting the representations of {$H$} to which the fields in the collection $\upphi$ belong to --
implies that if $\upphi$ is a solution, so is its gauge transform~$\upphi^\gamma$. 
In particular, if $\gamma$ is a compactly supported element of $\H$ whose support $D_\gamma \subset M$ is the ``hole", 
we~have that $\upphi=\upphi'$ on $M/D_\gamma$, while $\upphi \neq \upphi'$ on $D_\gamma$. 
It would then \mbox{appear} that the Cauchy problem is ill-posed, and the \mbox{theory} undeterministic. 

Then enters the 
 \emph{internal point-coincidence argument}: 
 The physical content of a gauge \mbox{theory} 
 is exhausted by the pointwise coincidental values of fields $\upphi$, 
whose \mbox{description} is $\H$-invariant. 
All solutions in the same \mbox{$\H$-orbit} $\O_\upphi$ thus represent the \emph{same} physical state.
\mbox{Deterministically} evolving 
physical d.o.f. 
are not represented by the individual mathematical fields $\upphi$, but in the $\H$-invariant \emph{relations} among them.

Together, these form a \emph{generalized hole argument} \mbox{arising} from the $\big(\!\Diff(M)\ltimes\H\big)$-covariance of the field \mbox{equation} of gRGFT. 
A \emph{generalized point-coincidence argument} answers it, implying that 
physical spatio-temporal and internal field d.o.f. are coextensive with the co-defining relations they participate in, and that the network of these relations \emph{define} physical spacetime \cite{JTF-Ravera2024c}.

\medskip

It is now clear that the boundary problem,  stated for $\Diff(M)$ and/or $\H$, has essentially the same logical \mbox{structure} as a hole argument. 
It commits  the same dual conceptual
\mbox{mistakes} of 1) considering mathematical fields $\upphi$ to \emph{directly \mbox{represent}} physical fields, and 2) considering $M$, and \emph{a fortiori}  boundaries $\d U$ of its bounded sets~$U$, as physical entities existing independently of the~fields. 
It~overlooks the relational resolution brought by the generalized point-coincidence argument: 
The boundary problem evaporates once it is recognized that  physical d.o.f. and physical spacetime, and its bounded regions, are \emph{relationally and invariantly defined}.  

These misconceptions, together with the various countermeasures put forward to solve the alleged issue (e.g. ``edge modes"),
would be avoided in a framework~where
both \emph{relationality} and \emph{strict invariance} are manifest. 
Such a reformulation 
of gRGFT would also give ready access to its physical observables. 
The challenge lies in how to technically implement this idea. This is where the DFM comes into play.

\section{Dressing Field Method: basics}\label{Dressing Field Method: basics}

Via the DFM \cite{Francois2023-a,JTF-Ravera2024gRGFT,JTF-Ravera2024-SUSY,JTF-Ravera2024ususyDFM,JTF-Ravera2024NRrelQM,JTF-Ravera2025DFMSusyReview&Misu,JTF-Ravera2025offshellsusy} one produces gauge-invariant variables out of the field space $\Phi=\{\upphi\}$ of a gRGFT. 
We~here  remind, and further expand on, the dressing procedure implemented in the DFM, both in the case of internal gauge symmetries $\mathcal{H}$ and in the case of  $\Diff(M)$. 
We shall also introduce the notion of \emph{dressed spacetime regions}, which will be pivotal in Section \ref{Dissolving the boundary problem in General Relativity}. 
The following is a  field-theoretical presentation, we refer to \cite{Francois2023-a,JTF-Ravera2024gRGFT} for a formulation in terms of differential bundle geometry.
\vspace{-3mm}

\subsection{The case of  gauge symmetries}
\label{The case of internal gauge symmetries}

Consider a GFT with field content $\upphi=\lbrace{ A, \varphi \rbrace}$, where $A$ is the 1-form gauge potential and $\varphi$ represents the matter fields content, both supporting the action of the gauge group $\mathcal{H}$ -- e.g. Yang-Mills theory and QED -- so that they
gauge-transform as
\begin{align}
\label{gaugetrAvphi}
    A^\gamma:=  \gamma\- A \gamma + \gamma\- d\gamma, \quad \varphi^\gamma:= \gamma\- \varphi, \quad \gamma \in \H .
\end{align}
A \emph{dressing field} is{, by definition,} 
a smooth map 
\begin{align}
\label{dressing-field-101}
    u : M \rightarrow H , \qquad \text{such that} \quad u^\gamma = \gamma\- u , \quad \gamma \in \H .
\end{align}
A key aspects of the DFM it that a dressing field should always be extracted from the field content of the theory. 
This means that it has to be a \emph{field-dependent dressing field} $u=u[\upphi]$, so that $u^\gamma := u[\upphi^\gamma] = \gamma\- u [\upphi]$.

When such a dressing field is found/built, we can define the \emph{dressed fields} $\upphi^u = \lbrace{ A^u , \varphi^u \rbrace}$, which are given by
\begin{align}
\label{dressed-fields}
A^u:=  u\- A u + u\- du, \quad \varphi^u:= u\- \varphi.
\end{align} 
This illustrates the DFM ``rule of thumb": To dress fields or functional thereof, we compute first their gauge transformations, then formally substitute the gauge parameter $\gamma$ with the dressing field $u$. 
The resulting expressions are $\H$-invariant by construction.
We stress, however, that since $u \notin \H$, dressed fields are not gauge transformed fields, and in particular a dressing via the DFM \emph{is not} a gauge-fixing \cite{Berghofer-Francois2024,JTF-Ravera2025DFMSusyReview&Misu}.

When $u=u[\upphi]$ is a field-dependent dressing fields, the DFM has a natural \emph{relational} interpretation \cite{JTF-Ravera2024c,JTF-Ravera2024gRGFT}: the dressed fields $\{A^{u[A, \vphi]}, \vphi^{u[A, \vphi]}\}$ in \eqref{dressed-fields} represent the \emph{gauge-invariant} \emph{relations} among the physical internal d.o.f. \mbox{embedded} in $\{A, \vphi\}$. 
They technically \mbox{implement} the internal point-coincidence argument.
Still, remark that these dressed field  
remain objects defined on the manifold $M$.

\smallskip

\paragraph{Dynamics}
The above pertain to the kinematics.
The dynamics of a GFT is specified by a Lagrangian $L=L(\upphi)$, that is typically required to be quasi-invariant under $\H$, i.e. 
$L(\upphi^\gamma)=L(\upphi) + db(\gamma;\upphi)$, for $\gamma \in \H$.
This guarantees the $\H$-covariance of the field equations $ E(\upphi)=0$ extracted from $L$. 

Given a dressing $u$, we  define the \emph{dressed \mbox{Lagrangian}} 
\begin{align}
\label{dressed-Lagrangian-int}
L(\upphi^u):=L(\upphi) + db(u;\upphi),
\end{align}
which is strictly $\H$-invariant by construction -- this is a case of the DFM rule of thumb. 
The field equations for the dressed fields, $ E(\upphi^u)=0$, have the \emph{same functional expression} as those of the ``bare" fields: here they {still differ by a boundary term, 
see \cite{JTF-Ravera2024gRGFT}.} 
The dressed field equations are \emph{deterministic}, they specify  uniquely the evolution of the physical {internal}  d.o.f. represented by the dressed fields \eqref{dressed-fields}.




\vspace{-2mm}

\subsection{The case of diffeomorphisms}\label{The case of diffeomorphisms}

Consider a general-relativistic theory with field content $\upphi=\{A, \vphi, g\}$, where $g$ is a metric field on $M$, supporting the pullback action of the group of diffeomorphisms, 
\begin{equation}
\label{Diff-trsf-fields}
\begin{aligned}
    & \upphi^\psi := \psi^* \upphi, \quad
    \psi \in \Diff(M),
    \\[1mm]
    & \text{i.e.} \quad \lbrace{ A^\psi, \varphi^\psi, g^\psi \rbrace} := \lbrace{ \psi^* A, \psi^* \varphi,\psi^* g \rbrace}.
\end{aligned}
\end{equation}
A dressing field for diffeomorphisms is a smooth map
\begin{align}
\label{diffeo-dressing-field}
    \upsilon: N \rightarrow M, \qquad \text{such that} \quad \upsilon^\psi := \psi\- \circ \upsilon, 
\end{align}
for any $\psi \in \Diff(M)${, $N$ being a model smooth manifold (s.t. dim$N=$dim$M$, typically $N=\mathbb{R}^n$).} 
As previously, a dressing field  should be extracted from the field content of the theory  
i.e. it should be a field-dependent dressing field $\upsilon = \upsilon[\upphi]$, so that $\upsilon^\psi := \upsilon (\psi^* \upphi) = \psi\- \circ \upsilon[\upphi]$.

Given such a dressing field $\upsilon$,  dressed fields  defined as
\begin{equation}
\label{diffeodressedfields}
\begin{aligned}
    & \upphi^\upsilon := \upsilon^* \upphi , \\[1mm]
    & \text{i.e.} \quad \lbrace{ A^\upsilon, \varphi^\upsilon, g^\upsilon \rbrace} = \lbrace{ \upsilon^* A, \upsilon^* \varphi,\upsilon^* g \rbrace},
\end{aligned}
\end{equation}
are $\Diff(M)$-invariant by construction. 
Remark again at play the DFM rule of thumb, here for \mbox{diffeomorphisms}.
For $\upsilon=\upsilon[\upphi]$, the dressed fields \eqref{diffeodressedfields} are manifestly \mbox{relational} variables: 
they represent $\Diff(M)$-invariant \mbox{\emph{relations}} among the physical spatio-temporal d.o.f. \mbox{embedded} in $\upphi$. 
They thus implement the first of the dual insights stemming from the point-coincidence \mbox{argument}.
Regarding the second, 
observe that they $\emph{are not}$ objects \mbox{defined} on the ``bare" manifold $M$.
\smallskip

\paragraph{Dressed regions}
The dressed fields \eqref{diffeodressedfields} live on  
field-dependent \emph{dressed regions} defined by
\begin{align}
\label{fielddepdrreg}
    U^\upsilon = U^{\upsilon[\upphi]} 
    := \upsilon[\upphi]\- (U),
\end{align}
with $\upsilon\-$ the inverse map of $\upsilon$, such that $\upsilon \circ \upsilon\-=\id_M$. 
Crucially, these are $\Diff(M)$-invariant:
Indeed,  defining the \emph{right action} of $\Diff(M)$ on subsets $U\subset M$ as $U \mapsto U^\psi:=\psi\- \circ U$ -- the action \eqref{Diff-trsf-fields} of $\Diff(M)$ on fields $\upphi$ being also a right action --  we find
\begin{equation}
\label{inv-dressed-regions}
\begin{aligned}
    (U^\upsilon)^\psi & = (\upsilon[\upphi]^\psi)\- \circ (U^\psi) \\
    & = \upsilon[\upphi]\- \circ \psi \circ \psi\- \circ (U) = U^\upsilon.
\end{aligned}
\end{equation}
The reason for the definition \eqref{fielddepdrreg}, and the justification of the claim that dressed fields $\upphi^\upsilon$ live on such regions, simply comes from integration theory: It tells us e.g. that 
the action is invariant,
\begin{align}
    S:=\int_U L(\upphi) 
    = \int_{\,\psi\-(U)}\!\!\!\!\!\!\!\!\!\!\!\!\psi^*L(\upphi)
    = \int_{\,U^\psi} \!\!\!\! L(\upphi^\psi)
    =S^\psi,
\end{align}
where $L(\upphi)$ is the Lagrangian providing the dynamics of a general-relativistic theory, required $\Diff(M)$-covariant $L(\upphi^\psi)=\psi^*L(\phi)$. 
By the DFM rule of thumb and \eqref{diffeodressedfields}, this gives
\begin{equation}
\label{int-dressed-fields}
S=\int_U L(\upphi) 
= \int_{\,\upsilon\-(U)}\!\!\!\!\!\!\!\!\!\!\!\! \upsilon^*L(\upphi)
=\int_{\,U^\upsilon}\!\!\!\! L(\upphi^\upsilon)
=:S^\upsilon.
\end{equation}  
The  $\Diff(M)$-invariant regions $U^{\upsilon[\upphi]}$ thus represent \emph{physical} regions of spatio-temporal events,  
a physical spatio-temporal event being a field-dependent $\Diff(M)$-invariant point $x^{\upsilon[\upphi]}:= \upsilon[\upphi]\-(x) \in U^\upsilon$,  
realizing the basic intuition behind the point-coincidence argument.

{We may call $M^\upsilon := \left\{ U^\upsilon\, |\, \forall\ U\subset M\right\}$ }
the \emph{manifold of physical spatio-temporal events}: we remark that it \emph{does not exist} \mbox{independently} from the fields $\upphi$. 
The~physical \emph{spacetime}, i.e. the physical Lorentzian manifold, is then  $(M^\upsilon,g^\upsilon)$. 
This  technically implements the second of the dual insights stemming from  the point-coincidence argument,  i.e. that physical spacetime 
is \emph{relationally defined}, in a $\Diff(M)$-invariant way, by its physical field content, and ``\emph{does not claim existence on its own,
but only as a structural quality of the field}."
\smallskip

\paragraph{Dynamics}
The Lagrangian being $\Diff(M)$-covariant, so are the field equations $E(\upphi)=0$ derived from it. 
Given a dressing $\upsilon$,  the \emph{dressed \mbox{Lagrangian}} 
\begin{align}
\label{dressed-Lagrangian-diffeo}
L(\upphi^\upsilon):=\upsilon^*L(\upphi)
\end{align}
is strictly $\Diff(M)$-invariant by construction -- again a case of the DFM rule of thumb. 
The field \mbox{equations} for the dressed fields, $ E(\upphi^u)=0$, 
have the \emph{same \mbox{functional} expression} as the ``bare" ones \cite{JTF-Ravera2024gRGFT}, but have a well-posed Cauchy problem,  uniquely
  determining  the \mbox{evolution} of the physical spatio-temporal  d.o.f. \mbox{represented} by the dressed fields \eqref{diffeodressedfields}.

\subsection{{Examples of dressing fields}}
\label{Examples of dressing fields}

{
One can find in the gRGFT literature many instances fleshing out the above general framework \cite{JTF-Ravera2024gRGFT}.
Let us  mention three basic examples illustrating group-valued dressing fields as defined in Section \ref{The case of internal gauge symmetries}.}

{
First, consider scalar electromagnetism, with $\upphi=\{A, \phi\}$ and $\H=\U(1)$: the $U(1)$-valued dressing field $u=u[\phi]:=e^{i\theta}$ is the phase of the $\CC$-scalar field $\phi=\rho e^{i\theta}$,   supporting indeed the defining $\U(1)$-gauge transformation $u^\gamma=u[\phi^\gamma]=\gamma\- u[\phi]$ for $\gamma \in \H=\U(1)$, so that $\upphi^u=\{A^u, \phi^u\}=\{A+id\theta, \rho\}$ are the $\U(1)$-invariant physical fields. 
This e.g. grounds attempts at providing invariant  account of the Aharonov-Bohm effect \cite{Wallace2014, Francois2018}, as well as invariant formulation of the Abelian Higgs model bypassing the notion of spontaneous symmetry breaking (SSB). 
See  \cite{Berghofer-et-al2023} section~5.3 (also \cite{Rubakov1999} chap.~6).
In spinorial electromagnetism, with $\upphi=\{A, \uppsi\}$ and $\H=\U(1)$, the extraction of a $\U(1)$-dressing field $u=u[A]$ from the (holonomy of the)  gauge potential yielding dressed fields $\upphi^u=\{A^u, \uppsi^u\}$ is the formal underpinning of Dirac's  proposal \cite{Dirac55, Dirac58} for an invariant quantization scheme in QED \footnote{In that context, in \cite{Dirac55} Dirac says of the dressed fermion field $\uppsi^{u[A]}:=u[A]\-\uppsi$ that, upon quantization, it is 
``the operator of creation of an electron \emph{together with its
Coulomb field}, or possibly the operator of absorption of a positron \emph{together with its Coulomb field}. 
It is to be contrasted with the operator $\uppsi$, which gives the creation or absorption of a bare particle.
\emph{A theory that works entirely with gauge-invariant operators has its electrons and positrons always
accompanied by Coulomb fields around them}, which is very reasonable from the physical point of
view.'' Emphasis is his.}.
}

{
Secondly, and similarly to the Abelian case, in the electroweak model a $\SU(2)$-dressing field $u=u[\varphi]$ may be extracted from the $\CC^2$-scalar field $\vphi$; the dressed gauge potential 
$A^{u[\vphi]}=\{\upgamma, W^\pm, Z^0 \}$, the dressed fermions  $\uppsi^{u[\vphi]}=\{ (\ell_L, \nu_\ell), \ell_R, (q_u, q_d) \}$ and  dressed scalar $\vphi^{u[\vphi]}=H$, are the $\SU(2)$-invariant fields existing in \emph{both} the massless and massive phase. See \cite{Francois2018, Attard-et-al2019} and \cite{Berghofer-et-al2023} section 5.4.
This grounds invariant approaches avoiding the notion of SSB: from Higgs \cite{Higgs66} and Kibble \cite{Kibble67} \footnote{Kibble wrote: ``It is perfectly possible to describe [the model] without ever
introducing the notion of symmetry breaking, merely by writing down the Lagrangian (66) [the dressed one $L(\upphi^u)$, eq. \eqref{dressed-Lagrangian-int}]. Indeed if the physical world were really described by this model, it is
(66) rather than (64) [the bare Lagrangian $L(\upphi)$] to which we should be led
by experiment.'' Kibble was publishing just eight months before the famous paper by Weinberg ``\emph{A model of leptons}'' \cite{Weinberg1967} containing the electroweak theory.}, to the so-called Fröhlich-Morchio-Strocchi (FMS) approach \cite{Frohlich-Morchio-Strocchi80, Frohlich-Morchio-Strocchi81, Maas2019}, which is in keeping with Elitzur theorem \cite{Elitzur1975} and indeed the basis of concrete lattice implementations of the Standard Model \cite{Creutz2024, Creutz2025}.}
 
{
Thirdly, consider gauge gravity coupled to an effectively described (i.e. non-spinorial) matter ``fluid" $\upvarphi$, with $\H=\SO(1,3)$ and  $\upphi=\{\t A, \upvarphi\}=\{A, e, \upvarphi\}$ where $A$ is the spin connection and $e={e^a}_\mu dx^{\,\mu}=:\b e \cdot dx$ is the tetrad 1-form, so that $\t A=A+e$ is a Poincaré-valued Cartan connection \cite{JTF-Ravera2024review, Francois2021}. 
A natural $\SO$-dressing field is none other than the tetrad field, $u=u[\t A]:=\b e$, which satisfies indeed $u^\gamma=u[\t A^\gamma]=\gamma\- \b e$ for $\gamma\in \SO(1,3)$~\footnote{Remark that here  $u[\t A]=\b e : M \rarrow GL(4) \supset SO(1,3)$, i.e. the dressing field takes value in a group $G \supset H$, yet still has the defining $\H$-transformation property. 
This more general situation is fully accounted for by the DFM \cite{GaugeInvCompFields, Francois2014, JTF-Ravera2024gRGFT}, of which a slightly simplified version was given in Section \ref{The case of internal gauge symmetries}.
Early on, \cite{GaugeInvCompFields} section 4.3  discussed  the possibility of extracting an $SO$-valued dressing field from  the $GL$-valued tetrad (via  an extension of the Schweinler-Wigner orthogonalization procedure \cite{Schweinler-Wigner}). See also  Appendix B of \cite{Francois2014} for  details, as well as footnote 12 in \cite{Francois2018}, and section 4.3.2. of \cite{Francois-et-al2021}. 
Such a ``minimal dressing'' extraction from the tetrad is the basis of a recent quantum gravity approach by Thiemann \cite{Thiemann2024}, and called there the ``triangular gauge''; an unfortunate choice of terminology given the clear   conceptual and mathematical distinction between dressing and gauge-fixing \cite{Berghofer-Francois2024}.
}. 
The dressed Cartan connection is $\t A^u=(A^u, e^u)=(\b e\- A \b e + \b e \- d\b e, \, \b e\- e) =:(\Gamma, dx)$, where $\Gamma$ is none other than the $\SO$-invariant, $\gl$-valued affine \mbox{connection}.
We~remark that the dressed $\eta$-transposed tetrad field $e^t= dx\cdot \b e^T \eta$, $\eta$ the Minkowski metric,  just gives the $\SO$-invariant metric  on $M$: $(e^t)^u:=e^t \b e = dx \cdot \b e^T \eta \b e =: dx\cdot \b g$, the metric field being $g=g_{\mu\nu}\, dx^{\,\mu}\!\otimes dx^\nu=:\b g\, dx\!\otimes\!dx$. 
The bare Lagrangian $L(A, e, \upvarphi)$ is dressed as $L(\Gamma, g, \upvarphi)$; 
the DFM here then  accounts for the  switch between the tetrad  and metric  formulations of gravity. 
This case we revisit in Section \ref{Scalar coordinatization of GR via the DFM}  to illustrate dressings for diffeomorphisms and, thereby, the following core message.  
}

\subsection{No boundary problem in gRGFT}
\label{No boundary problem in gRGFT}

Gathering the results of  
{Sections \ref{The case of internal gauge symmetries} and \ref{The case of diffeomorphisms},}
we \mbox{obtain} a  $\big(\!\Diff(M)\ltimes \H)$-invariant and manifestly \mbox{relational} \mbox{reformulation} of gRGFT:
Defining a \emph{complete} field-dependent dressing field as the pair  $(\upsilon, u)=(\upsilon[\upphi],  u[\upphi])$, we define, using \eqref{dressed-fields}-\eqref{diffeodressedfields}, the fully invariant dressed fields  
\begin{align}
    \upphi^{(\upsilon, u)} := \upsilon^*(\upphi^u),
\end{align}
representing the physical spatio-temporal and internal field d.o.f. and their
 co-defining (coextensive) relations. 
As above, the fields $\upphi^{(\upsilon, u)}$ live on/define the physical manifold of spatiotemporal events $M^{(\upsilon, u)}= M^\upsilon$, and regions $U^\upsilon$ thereof \footnote{
The dressing field $(\upsilon,u)$ can be seen as the local version of a field-dependent dressing field $\bs u$ on a principal fiber bundle $P$ over $M$. The DFM procedure in this case allows to define the relational, physical `enriched' spacetime as a dressed bundle \cite{JTF-Ravera2024c,BFR-fbsubst,JTF-Ravera2024gRGFT}.}.
Their dynamics is given by the \mbox{invariant} dressed Lagrangian $L( \upphi^{(\upsilon, u)}):= \upsilon^*L(\phi^u)$, 
from which derive the field \mbox{equations} $E( \upphi^{(\upsilon, u)})=0$ with a well-posed Cauchy problem. 
This implements both dual aspects of the generalised point-coincidence argument.
\smallskip

An immediate consequence of all the above is that a physical, relationally-defined, boundary $\partial U^\upsilon$ is by \mbox{definition} $\big(\!\Diff(M)\ltimes \H\big)$-invariant.
This dissolves the boundary problem, understood as the claim that
``$\Diff(M)$ and/or
$\H$ symmetries are broken at \emph{spacetime boundaries}”. 
If it is meant literally, it is wrong. 
If it is meant as ``fields $\upphi$, or functional thereof, defined
at $\d U$ are not $\big(\!\Diff(M)\ltimes \H\big)$-invariant" or as ``$\d U$ is not preserved by $\Diff(M)$", it is technically true but trivial and physically inconsequential.



\section{Dissolving the boundary problem in General Relativity}\label{Dissolving the boundary problem in General Relativity}

As should be clear from what precedes, 
the boundary problem for GFT is easier to tackle/dissolve: it is sufficient to build $\H$-invariant field variables $\upphi^u$, since the manifold $M$ on which the fields $\upphi$ are defined does not transform under the action of $\H$  \footnote{The origin of the latter being indeed the group $\text{Aut}_v(P)$ of vertical automorphisms  of the principal fiber bundle $P$ over $M$, whose geometry underlies the kinematics of GFT. These automorphisms of $P$ induce, by definition,  the identity transformation $\id_M$ of $M$. They form a  (normal) subgroup of the full group $\text{Aut}(P)$ of automorphisms of $P$, which induces diffeomorphisms $\Diff(M)$ of~$M$. The geometry of principal bundles is the unifying geometric framework for gRGFT. See \cite{JTF-Ravera2024gRGFT, JTF-Ravera2024c}.}.
The \mbox{physical} configuration of internal d.o.f. represented by $\upphi^u$ is  $\H$-invariant across $M$, in particular at $\d U$ for $U\subset M$.
For applications of the framework laid in Section \ref{The case of internal gauge symmetries}, e.g. to the electroweak model, to electromagnetism coupled to scalar fields, and other GFTs, see  \cite{Berghofer-et-al2023,JTF-Ravera2024gRGFT}.


The boundary problem in GR physics is more involved since the manifold $M$ itself transforms under~$\Diff(M)$. 
The dressing field $\upsilon[\upphi]$ is used to  dress both the bare fields $\upphi$ of the theory and  (regions $U$ of)~$M$. 
The framework of Section \ref{The case of diffeomorphisms} encompasses diverse \mbox{variants} of  ``scalar coordinatization" in GR. 
For example, 
in the approach à la Kretschmann–Komar{-Bergmann} \cite{Komar:1958ymq, Bergmann-Komar1960, Bergmann1961,  Bergmann:1972ud} for \mbox{vacuum} GR, a dressing field is extracted from the bare metric, $\upsilon=\upsilon[g]$ (otherwise seen as a a ``$g$-dependent coordinate system"); 
the dressed metric $g^\upsilon:=\upsilon[g]^*g$ is  ``self-dressed", it represents the invariant structure among  the d.o.f. of the physical  gravitational field. 
In \mbox{approaches} à la {DeWitt \cite{DeWitt1960}, or} Brown–Kucha\v{r} \cite{Brown:1994py, Rovelli:2001my},
if matter is \mbox{described} effectively as a fluid (gas, particles, dust, etc.), it provides scalars $\upvarphi=\upvarphi^a$, $a\in \{0, \ldots, n-1\}$ 
from which one gets the dressing field $\upsilon[\upvarphi]$;
the dressed metric $g^\upsilon:=\upsilon[\upvarphi]^*g$  represents~the \mbox{invariant} relational structure between the d.o.f. of the metric and of the effective matter field --  otherwise seen as the metric ``written" in the physical matter frame {\footnote{{DeWitt \cite{DeWitt1960} would say that the dressing field $\upsilon[\upphi]$ provides what he calls ``intrinsic coordinates'', whereby one may build invariants: in our language, the dressed fields $\upphi^\upsilon:=\upsilon^*\upphi$. To~quote him more extensively, he argues that ``what appears to be needed, if one insists on maintaining manifest covariance [we would rather say, $\Diff(M)$-invariance], is a mean of constructing \emph{local invariants}. A possible procedure is to introduce four independent scalars $\zeta^A$, $A=0,1,2,3$ [yielding a dressing field for $\Diff(M)$ as defined in eq.  \eqref{diffeo-dressing-field}], formed out of the metric and its derivative [i.e. a dressing field $\upsilon[g]$] and then to use these to define an \emph{intrinsic coordinate system}.'' His emphasis. 
DeWitt cites Komar and Bergmann \cite{Komar:1958ymq, Bergmann-Komar1960} as doing just that, 
but noting possible difficulties with using the metric; to avoid them he  continues ``[...] we shall introduce directly into the discussion an additional physical system. 
This system will serve to furnish us with a reasonably full-proof set of intrinsic coordinates [i.e. a dressing field $\upsilon[\upphi]$] while at the same time forming a \emph{combined physical system} with the the gravitational field.'' 
The emphasis here is ours, and highlights the \emph{relational} character of the physical system to be analyzed.
He continues ``In principle, any additional system which provides a `useful' set of four scalars will do.'' -- see indeed Section IV.B for an illustration -- adding 
``It might be supposed that [the additional system] has merely a technical utility, constituting an otherwise foreign element in the discussion. Such is by no means the case. The role played by the medium in providing a \emph{physical coordinate system} proves to be a fundamental one [...]'' (our emphasis again), with which we fully agree.  Echoing (perhaps unwittingly) Einstein's point-coincidence argument described earlier, DeWitt indeed insists that ``A [...] medium of some kind is needed [...] in order to give an operational meaning to the concept of `space-time geometry' in the first place.''
We thank a referee for pointing out this reference by DeWitt.}}.}
In both cases, even though it is not usually done, other \mbox{physical} matter and/or interaction fields may be described as $\{A^\upsilon,\varphi^\upsilon\}$,
while physical regions of events  $U^\upsilon \subset M^\upsilon$ are  defined either via matter $\upsilon[\vphi]$, or the gravitational field $\upsilon[g]$.

\smallskip

In the following, we illustrate the construction for the case of GR:  assuming that a dressing field $\upsilon$ for $\Diff(M)$ has been built from the field content $\upphi$  of the theory, 
we want to prove the invariance of the dressed metric, i.e. of the physical gravitational field. 
We shall do so twice over, to showcase both the abstract and computationally concrete versions. 
For this we need to go over the basics.

\subsection{Physical boundaries in relational GR}\label{Physical boundaries in relational GR}

Recall that  diffeomorphisms $\psi \in \Diff(M)$ are smooth maps $\psi: M \to M$, $x' \mapsto \psi(x')$, with smooth inverse $x \mapsto \psi\-(x)$. 
Their linearization yields  vector fields $X := \tfrac{d}{d\tau}\psi_\tau\big|_{\tau=0}$, with $\psi_{\tau=0}=\id_M$,  which constitute the Lie algebra $\diff(M)\simeq \Gamma(TM)$. 
As previously mentioned, the right action of $\Diff(M)$ on regions $U\subseteq M$ is defined as  is $U \mapsto \psi\-(U)$. 
The group $\Diff(M)$ acts on vector fields by \emph{pushforward}, $\psi_*: TM \to TM$, ${\mathfrak X}_{|x} \mapsto (\psi_*{\mathfrak X})_{|\psi(x)}$.
Its action  on differential forms, and covariant tensors, defines the  \emph{pullback}, $\psi^*: T^*M \to T^*M$, $\alpha_{|x}\mapsto (\psi^*\alpha)_{|\psi^{-1}(x)}$.
The latter is a also a right action. 
The  linearisation of these actions defines the Lie derivative: 
On vector fields, $\L_X \mathfrak X := \tfrac{d}{d\tau} \psi_{\tau*} {\mathfrak X} \big|_{\tau=0} =[X, \mathfrak X]$.  
On differential forms and covariant tensors, $\L_X \alpha := \tfrac{d}{d\tau} \psi^*_\tau \,\alpha\,\big|_{\tau=0}$. 

The $\Diff(M)$-transformation of the metric, a \mbox{symmetric} covariant 2-tensor, is thus 
$g_{|x} \mapsto(\psi^*g)_{|\psi^*(x)}$ or  $g_{|\psi\- (x)} \mapsto (\psi^* g)_{|x}$.
And linearly, $\L_X g = \tfrac{d}{d\tau}\, \psi^*_\tau g\, \big|_{\tau=0}$.
We introduce the following compact matrix notation:
\begin{align}
    g = \b g \, dx \otimes dx , \quad \mathfrak X= \b{\mathfrak X} \cdot \partial_x , 
\end{align} 
where $\b g$ and $\b{\mathfrak X}$ are matrix-valued 0-forms on $M$, i.e. the `coordinates representatives' of $g$ and $\mathfrak X$, and \mbox{$dx(\d_x)=1$}.
 Explicitly, in components this is  $g=g_{\mu \nu} \, dx^{\mu} \otimes dx^{\nu}$ and  $\mathfrak X={\mathfrak X}^\mu \d_\mu$, 
 i.e.  $\b g=g_{\mu \nu}$ and $\b {\mathfrak X}=\mathfrak{X}^\mu$, and $dx^{\,\mu}(\d_\nu)=\delta^\mu_\nu$, with $\mu,\nu\in \{0,1,\ldots,n-1\}$. 
We have then by definition of the metric that
$g_{|x}({\mathfrak X}_{|x}, {\mathfrak Y}_{|x}) = \b{\mathfrak X}_{|x}^T\, \b g_{|x} \, \b{\mathfrak Y}_{|x}$ is a number, 
with $\b{\mathfrak X}^T$  the matrix transpose of the column matrix $\b{\mathfrak X}$.
 
We have the expression for the pushforward,
\begin{align}
\label{pushforward-X}
     \psi_* {\mathfrak X} = (\overline{\psi_* {\mathfrak X}})_{|x} \cdot \partial_x = G (\psi)_{|x} \b {\mathfrak X}_{|x} \cdot \partial_x ,
\end{align}
where $G(\psi):=\tfrac{\d\psi}{\d x}$ denotes the Jacobian matrix of $\psi$. 
The fundamental duality relation between pullback and pushforward is, applied to $g$,
\begin{align}
\label{duality-pullback-pushforward}
    (\psi^* g)_{|x} ({\mathfrak X}_{|x},{\mathfrak Y}_{|x}) = g_{|\psi(x)} (\psi_* {\mathfrak X}_{|\psi(x)},\psi_* {\mathfrak Y}_{|\psi(x)}) ,
\end{align}
which yields the $\Diff(M)$-transformation  of   $\b g$:
\begin{align}
\label{diff-trsf-comp-g}
    (\overline{\psi^* g})_{|x}  = G(\psi)^T_{|x}\, [\,\b g \circ \psi(x)\,]\, G(\psi)_{|x},
\end{align}
from which we can find its Lie derivative,
    \begin{align}
        \L_X \b g_{|x}  :\!&= \tfrac{d}{d\tau} (\overline{\psi^*_\tau g})_{|x} \big|_{\tau=0} \\[1mm]
        & = \left(\tfrac{d}{d\tau} G(\psi_\tau)^T_{|x} \big|_{\tau=0} \right) [\,\b g \circ \psi_{\tau=0}(x)\,] \,G(\psi_{\tau=0})_{|x} \notag\\[1mm]
        & \quad+ G(\psi_{\tau=0})^T_{|x} \,[\,\b g \circ \psi_{\tau=0}(x)\,] \left( \tfrac{d}{d\tau} G(\psi_\tau)_{|x}\big|_{\tau=0}\right) \notag\\[1mm]
        & \quad + G(\psi_{\tau=0})^T_{|x} \underbrace{\tfrac{d}{d\tau} \,[\,\b g \circ \psi_\tau(x)\,]\big|_{\tau=0}}_{=:X(\b g)=\iota_X d \b g} G(\psi_{\tau=0})_{|x} \, , \notag
    \end{align}
where $\iota_X$ denotes the contraction operator on forms. 
\mbox{Introducing} the linearized Jacobian
\begin{align}
    J(X)_{|x} := \tfrac{d}{d\tau} G(\psi_{\tau})_{|x}\big|_{\tau=0}, 
\end{align}
we  finally obtain the known expression
\begin{align}
    \L_X \b g = \iota_X d \b g + J(X)^T \b g + \b g J(X) .
\end{align}
This recap was necessary to fully understand how the DFM ought to apply in concrete computations.
\medskip

Given a dressing field $\upsilon=\upsilon[\upphi]$ for diffeomorphisms, 
the proof of the invariance of dressed points $x^\upsilon:=\upsilon\- (x)$ and  bounded regions $U^\upsilon:= \upsilon\-(U)$ with  boundaries $\d U^\upsilon$ of the physical manifold of events is already given by \eqref{inv-dressed-regions}. 
As  already mentioned, the \emph{physical spacetime} is  $(M^\upsilon,g^\upsilon)$, with $g^\upsilon:= \upsilon^*g$ the dressed metric field. 
We need only to prove its $\Diff(M)$-invariance to definitely establish
that there is no boundary problem.



\smallskip

\paragraph{Abstract proof} 
The intrinsic, abstract, proof of the $\Diff(M)$-invariance of the dressed metric $g^\upsilon$ is easy,
\begin{equation}
\begin{aligned}
    (g^\upsilon)^\psi & = (\upsilon^\psi)^* g^\psi = (\psi\- \circ \upsilon)^* (\psi^* g) \\
    & = \upsilon^* {\psi\-}^* \psi^* g = \upsilon^* g =: g^\upsilon .
\end{aligned}
\end{equation}
It is correspondingly trivial to prove the vanishing of its Lie derivative:
\begin{align}
    \L_X g^\upsilon 
    :=
    \tfrac{d}{d\tau} \,(g^\upsilon)^{\psi_\tau}\big|_{\tau=0}
    =
     \tfrac{d}{d\tau}\,  (g^\upsilon)\big|_{\tau=0}
    \equiv 0 .
\end{align}
Remark that the spirit of the proof applies to any tensor or pseudo-tensor.

\smallskip

\paragraph{Computational proof}
The previous results are not surprising, conceptually: as we have stressed, the dressed metric $g^\upsilon$ does not ``live" on $M$, but on the physical manifold $M^\upsilon$. 
The dressing field being a smooth map, 
$\upsilon: M^\upsilon \rarrow M$, $y:=x^\upsilon \mapsto \upsilon(y)=x$, 
its pushforward is $\upsilon_*: TM^\upsilon \rarrow TM$, 
\begin{align}
\label{pushforward-upsilon}
  {\mathcal X} \mapsto    \upsilon_*  {\mathcal X} = (\overline{\upsilon_* {\mathcal X}})_{|y} \cdot \partial_y = G (\upsilon)_{|y} \b {\mathcal X}_{|y} \cdot \partial_y,
\end{align}
where $G(\upsilon):=\tfrac{\d\upsilon}{\d y}$ is the Jacobian matrix of the dressing field. We have then the pullback/pushforward duality relation between the bare and dressed metrics,
\begin{align}
\label{duality-pullback-pushforward-dr}
    (\upsilon^* g)_{|y} ({\mathcal X}_{|y},{\mathcal Y}_{|y}) = g_{|\upsilon(y)} (\upsilon_* {\mathcal X}_{|\upsilon(y)},\upsilon_* {\mathcal Y}_{|\upsilon(y)}) ,
\end{align}
which yields the component expression of the dressed metric in terms of that of the bare metric:
\begin{align}
\label{dressed-metric-comp}
    \overline{g^\upsilon}_{|y} = \overline{\upsilon^* g}_{|y} = G(\upsilon)^T_{|y} \,[\, \b g \circ \upsilon (y)\,]\, G(\upsilon)_{|y}.
\end{align}
Again, this can be understood as the components of the physical metric/gravitational field as determined by the reference frame $\upsilon[\upphi]$ provided by the other fields $\upphi$. 

\noindent
Now, the $\Diff(M)$-transformation of the Jacobian of $\upsilon$ is
\begin{equation}
\label{trjac}
\begin{aligned}
    G(\upsilon)^\psi := G(\upsilon^\psi) = G(\psi\- \circ \upsilon) 
    & = G(\psi\-) \, G(\upsilon) \\
    & = G(\psi)\- G(\upsilon),
\end{aligned}
\end{equation}
where in the last step we use the well-known property that the Jacobian of the inverse of a map is the inverse of the Jacobian of the map \footnote{
Remark that, seeing $G(\upsilon): M^\upsilon \rarrow GL(n)$, this transformation closely resembles that of an `internal' dressing \eqref{dressing-field-101}.}.
Using \eqref{diff-trsf-comp-g} and \eqref{trjac}, we can check explicitly  the invariance of $\overline{g^\upsilon}$:
    \begin{align}
        \overline{g^\upsilon}^{\,\psi} 
        & =
        {G(\upsilon)^\psi}^T [\,\b g^\psi \circ \upsilon^\psi\,]\, G(\upsilon)^\psi  \notag\\
        & = G(\upsilon)^T {G(\psi)\-}^T [G(\psi)^T (\b g \circ \psi \circ \psi\- \circ \upsilon ) G(\psi)]  \notag\\
        & \qquad  \cdot G(\psi)\-G(\upsilon) \notag \\
        & = G(\upsilon)^T \, [\, \b g\circ \upsilon \,]\, G(\upsilon) =: \overline{g^\upsilon}. 
    \end{align}
Correspondingly, using the linear version of \eqref{trjac},
\begin{align}
    \L_X G(\upsilon) = \tfrac{d}{d\tau} G(\upsilon)^{\psi_\tau}\big|_{\tau=0} = - J(X)\, G(\upsilon) ,
\end{align}
we may explicitly compute:
\begin{equation}
\begin{aligned}
    \L_X \, \overline{g^\upsilon} 
    & = \L_X \left(G(\upsilon)^T[\b g \circ \upsilon] G(\upsilon)\right) \\
    & = 
    \left( \L_X G(\upsilon)^T \right)[\b g \circ \upsilon] G(\upsilon) \\
    &\quad + G(\upsilon)^T [\b g \circ \upsilon] \left( \L_X G(\upsilon) \right) + G(\upsilon)^T [\, \L_X \b g_{|x}\,] G(\upsilon) \\
    &\quad  + G(\upsilon)^T \tfrac{d}{d\tau}\, \left[\, \b g \circ \psi\-_\tau (x) \big|_{\tau=0}\, \right]\,G(\upsilon) \\
    & = - G(\upsilon)^T J(X)^T [\,\b g \circ \upsilon\, ]\, G(\upsilon) \\
    &\quad  + G(\upsilon)^T [\,\b g \circ \upsilon\,] (-J(\upsilon)G(\upsilon)) \\
    &\quad  + G(\upsilon)^T \lbrace{ \iota_X d \b g + J(X)^T \b g + \b g J(X) \rbrace}\, G(\upsilon) \\
    &\quad  + G(\upsilon)^T [\underbrace{-X(\b g)}_{-\iota_X d \b g}\,]\, G(\upsilon) \equiv 0.
\end{aligned}
\end{equation}
Again, the principle of the proof applies to any tensor or pseudo-tensor.

\subsection{Scalar coordinatization of GR via the DFM}\label{Scalar coordinatization of GR via the DFM}

We now illustrate the above to the case of 
 \mbox{$n=4$} GR, with cosmological constant $\Lambda$,
 and matter described \mbox{phenomenologically} as a {(perfect)} fluid {\footnote{{We may e.g. quote DeWitt \cite{DeWitt1960} again; if ``[i]n principle, any additional system which provides a ‘useful’ set
of four scalars [dressing $\upsilon[\upphi]$] will do. Actually, we shall choose the most intuitively obvious system possible, namely, a stiff elastic medium carrying a framework of clocks.'' whose ``physical constitution'' is considered ``only phenomenologically''.}};
}
 {it supplies a set of scalar fields $\upvarphi=\upvarphi^a : U\subseteq M \rarrow N=\RR^4$, 
 $a=\{1,\ldots ,4\}$, characterizing the matter/fluid distribution and entering the expression of its covariantly conserved stress-energy tensor $\mathsf{T}=\mathsf{T}(g, \upvarphi)$,  $\nabla^g \mathsf T=0$, itself derivable as the Hilbert stress-energy tensor of an effective Lagrangian $L_\text{\tiny{matter}}(g, \upvarphi)$ -- whose precise form is not needed here {\footnote{The scalars could be {taken to be, if not} as the \emph{components} of the 4-velocity of the fluid particles $\upvarphi^a=u^a$ {(as determined by any arbitrary \emph{frame field}, i.e. a  section of the frame bundle $LM$ of $M$), at least  as} a (sub)set of scalars in terms of which the fluid  4-velocity is expressed, $u^a=u^a(\upvarphi)$, as is done in the so-called ``velocity-potential'' representations.
 A typical choice of effective Lagrangian is $L_\text{\tiny{matter}}(g, \upvarphi)= \rho\, \text{vol}_{g}$, with $\text{vol}_{g}$ the volume form induced by the metric field $g$ and $\rho=\rho(\upvarphi, \ldots)$ the rest energy density of the fluid expressed as a function of the  scalars $\upvarphi$ and possibly of other thermodynamical parameters (entropy per baryon, chemical potential, etc.). 
The field equations for the  scalars are equivalent to particle (baryon) number conservation and covariant conservation of the stress-energy tensor, $\nabla \mathsf{T}(g. \upvarphi)=0$. 
 Such Lagrangian description was pioneered notably by Taub \cite{Taub1954, Taub1969}, Schutz \cite{Schutz1970}, as well as  Carter \cite{Carter1973},  Kijowski  and~Tulczyjew\cite{Kijowski-Tulczyjew1979} -- see also  Brown \cite{Brown1993, Brown-Marolf1996} -- and is an integral part of the field of relativistic fluid dynamics and numerical relativity \cite{Andersson-Comer2007, Rezzolla-Zanotti2013}.
 {In this field,  it is also often the case that the fluid distribution is described by a set of scalars fields labelling fluid particles and  called ``Lagrangian coordinate fields", or yet ``comoving coordinates" (standard, unphysical, \mbox{coordinates} on $M$ being called  ``Eulerian coordinates"); these are clearly fit for our purpose (e.g. DeWitt \cite{DeWitt1960} uses just this viewpoint). 
 Velocity-potentials, or a subset thereof, are sometimes used as Lagrangian coordinates.}
 Keen~readers may detect  in this literature  many instances of the DFM~\mbox{philosophy}; 
  e.g. the reference manifold $N$ (that is $R^4$ in the case at hand), the source space of the dressing field $\upsilon$, generalizes what is variously called ``material space'' \cite{Kijowski-Tulczyjew1979}, ``fluid space'' \cite{Brown1993}, and ``matter space''  \cite{Brown-Marolf1996, Andersson-Comer2007} -- or yet ``fleet'' (of fluid particles) \cite{Brown-Marolf1996}. Also, dressed fields $\upphi^\upsilon$ extend what Carter calls ``material tensors'' \cite{Carter1973}, while $g^\upsilon$ relates e.g. to the ``matter space/fleet metric'' of \cite{Brown-Marolf1996}.}.}}
 {The~fields   considered  $\upphi=\{g, \upvarphi\}$  $\Diff(M)$-transform as}
\begin{align}
  \upphi^\psi
  =
  \{\psi^*g , \, \psi^*\upvarphi \}
  =
  \{ \psi^*g, \,\upvarphi \circ \psi \},
\end{align}
{and} the  Lagrangian of the theory is 
\begin{equation}
\label{lagrangian-GR-scalar}
\begin{aligned}
L_\text{{\tiny GR}}(\upphi) &= L_\text{{\tiny GR}}(g, \upvarphi) \\[1mm]
& = \tfrac{1}{2\kappa }\text{vol}_{g}  \big(\mathsf R(g) - 2 \Lambda \big) + L_\text{\tiny{matter}}(g, \upvarphi), \end{aligned}
\end{equation}
where $\kappa=\tfrac{8 \pi G}{c^4}$ is the gravitational coupling constant,  $\text{vol}_{g}$ is the volume 4-form induced by $g$, and $\mathsf R(g)$ is the  Ricci scalar.
{The Lagrangian $L_\text{{\tiny GR}}(\upphi)$} is $\Diff(M)$-covariant, and so are the  Einstein equations derived from it: 
$E(\upphi)= \mathsf{G}(g) +\Lambda g  - \kappa \mathsf{T} (g, \upvarphi)=0$, 
where $\mathsf{G}(g)$ is the Einstein tensor. 
{Remark that $\nabla^g\mathsf{T}=0$ implies that the fluid particles are in geodesic motion either if the pressure gradient vanishes, or if pressure itself does, in which case the fluid is a dust field (and, as said earlier, we then make contact with \cite{Brown:1994py, Rovelli:2001my}).}

We can identify a $\Diff(M)$-dressing field as 
\begin{align}
\upsilon=\upsilon[\upvarphi]:=\upvarphi^{-1}: \RR^4 \rarrow M.
\end{align}
As indeed, $\upsilon^\psi=\upsilon[\upvarphi^\psi]=\psi\- \circ \upsilon[\upvarphi]$. It allows to define dressed regions
\begin{align}
    U^\upsilon := \upsilon\- (U), \quad \text{such that} \quad (U^\upsilon)^\psi = U^\upsilon.
\end{align}
This gives a $\Diff(M)$-invariant relational definition of physical regions of events via ({the scalar distribution of) matter} 
as a \emph{physical} reference  system, as is expected from the point-coincidence argument. 
The dressed fields are   
\begin{align}
     \upphi^\upsilon
     =
     \{g^\upsilon,\,  \upvarphi^\upsilon\} 
     =
     \{\upsilon^* g ,\,  \id_{U^\upsilon} \}.
\end{align}
Here $\upvarphi^{\upsilon}:=$ {$\upsilon^* \upvarphi= \upvarphi \circ \upsilon = $} $\id_{U^\upsilon
}${, the matter distribution being ``self-dressed'',
just expresses the fact that the (values of) the scalars now \emph{are} the coordinates -- a.k.a.  \emph{Lagrangian} or \emph{comoving} coordinates (see footnote [68]).}
The $\Diff(M)$-invariant dressed metric $g^{\upsilon}$, {encoding} the geometric properties of $M^{\upsilon}$,   can {then} be understood as the physical gravitational field as measured in the coordinate system supplied by the matter distribution $\upvarphi$. 
In~abstract index notation, \eqref{dressed-metric-comp} gives
\begin{align}
\label{dressed-metric-index}
g^\upsilon_{ab} = {G(\upsilon)_a}^\mu \,  g_{\mu \nu} \, {G(\upsilon)^\nu}_b ,
\end{align}
with the Jacobian 
$G(\upsilon)=G(\upvarphi\-)=G(\upvarphi)\-\!=\left( \tfrac{\partial \upvarphi^a}{\partial x^\mu} \right)\-\!$.

The  Lagrangian of the dressed, relational, theory is 
\begin{align}
\label{Dressed-lagrangian-GR-scalar}
L_\text{{\tiny GR}}(g^{\upsilon}, \upvarphi^{\upsilon}) 
:\!&=
{\upsilon}^*L_\text{{\tiny GR}}(g, \upvarphi)
 \\[1mm]
& = \tfrac{1}{2\kappa }\text{vol}_{g^{\upsilon}}  \big(\mathsf R(g^{\upsilon}) - 2 \Lambda \big)  +   L_\text{\tiny{matter}}(g^{\upsilon}, \upvarphi^{\upsilon}). \notag
\end{align}
From 
it derive the
\emph{relational Einstein equations}
\begin{align}
\label{dressed-Einstein-eq}
    E(\upphi^\upsilon)=\mathsf{G}(g^{\upsilon}) +\Lambda g^{\upsilon} - \kappa \mathsf{T}(g^{\upsilon}, \upvarphi^{\upsilon})=0 ,
\end{align}
{with  $\mathsf{T} (g^\upsilon, \vphi^\upsilon)=:\mathsf{T}^\upsilon$ the  conserved dressed stress-energy tensor, $\nabla^{g^\upsilon} T^\upsilon=0$ (controlling the dynamics of $\vphi^\upsilon$).}
These are strictly $\Diff(M)$-invariant, and have a well-posed Cauchy problem \footnote{As stressed already above, the dressed field equations have the same functional expression as the bare ones. In components, they are related by
\begin{equation*}
\begin{aligned}
   \phantom{blabl} & \mathsf{G}^{\upsilon}_{ab} +\Lambda g^{\upsilon}_{ab} - \kappa \mathsf{T}^{\upsilon}_{ab} \\[1mm]
    &\qquad \qquad \quad  = {G(\upsilon)_a}^\mu \left( \mathsf{G}_{\mu \nu} + \Lambda  g_{\mu \nu} - \kappa  \mathsf{T}_{\mu \nu} \right) {G(\upsilon)^\nu}_b = 0.
\end{aligned}
\end{equation*}
Which superficially resembles the general-covariance of Einstein's equation, yet is conceptually distinct. 
}. 
Notice that, contrary to the bare equations, there is no more ``pure metric side" in the dressed field equations \eqref{dressed-Einstein-eq};  all its terms involve \emph{both} the physical metric and matter d.o.f. 
These are the equations that are confronted to experimental tests
\cite{JTF-Ravera2024gRGFT} 
{--
see e.g. \cite{JTF-RaveraDarkMatter&DFM2025} for an application to galaxy rotation curves analysis.}

\section{Conclusions}\label{Conclusions}


We have addressed the claim, often encountered in general-relativistic and gauge field theoretic physics, that 
``diffeomorphism symmetry, $\Diff(M)$, and/or  gauge \mbox{symmetries}, $\H$, are broken at spacetime boundaries", which we refer to as the boundary problem.
We have demonstrated this claim to have the same conceptual structure as a hole-type argument, and is thus defused by the point-coincidence argument.
Upon recognizing  physical (spacetime) regions and boundaries as \mbox{invariant} structures, relationally defined by the physical fields, the boundary problem conceptually \mbox{dissolves}.

This resolution was technically realized  through a \mbox{manifestly} invariant and relational reformulation of general-relativistic gauge field theories via the Dressing Field Method, whereby the point-coincidence argument is implemented automatically  by \mbox{defining} dressed fields and regions. 
We have illustrated our framework with a concrete application to the general-relativistic case, where the $\Diff(M)$-invariance of the dressed metric and spacetime regions is proven twice over, abstractly and in a concrete computational way, 
 showing that no boundary problem can arise technically. 

A related consequence of our framework is that it  \mbox{dissolves}~one aspect of the multifaceted ``problem of time", i.e. the worry that, since in classical GR proper time evolution on $M$ can be seen as a special case of \mbox{diffeomorphisms}, ``time evolution is `pure gauge' in general-relativistic physics", so that physical observables have no dynamics. 
This is otherwise known as the ``frozen formalism problem" \cite{Isham1992, Kuchar2011, AndersonE2012}. 
The issue arises only if one overlooks that $M$ \emph{is not}  the true physical spacetime. 
This~aspect of the problem of time  dissolves in a way \mbox{analogous} to the boundary problem, on account of the relational core of general-relativistic physics.
{
The same ``problem'' arises in  parametrized Mechanics (see e.g.  \cite{Rovelli2004, Rovelli-Vidotto2014}) and is likewise resolved relationally;
 we give the detailed treatment via DFM in \cite{JTF-Ravera-MFT}, leading to what we call \emph{Relational Quantization} -- see also \cite{JTF-Ravera2024NRrelQM}.}

\vspace{-3.7mm}

\begin{acknowledgments}
\vspace{-1mm}
J.F. is supported by the Austrian Science Fund (FWF), grant \mbox{[P 36542]}, 
and by the Czech Science Foundation (GAČR), grant GA24-10887S.
L.R. is supported by the 
GrIFOS research project, funded by the Ministry of University and Research (MUR, Ministero dell'Università e della Ricerca, Italy), PNRR Young Researchers funding program, MSCA Seal of Excellence (SoE), 
CUP E13C24003600006, ID SOE2024$\_$0000103, of which this paper is part.
\end{acknowledgments}

\section*{Author Contribution declaration}

Both authors share equal credit for the work done in this paper: J.F. and L.R. conceived of the presented idea, J.F. and L.R. developed the theoretical formalism, J.F. and L.R. performed all the calculations. Both J.F. and L.R. authors contributed equally to the writing of the manuscript.




\bibliography{bibliobdy.bib}

\end{document}